\documentclass{mem}
\usepackage{natbib}
\usepackage{txfonts}
\usepackage{balance}
\usepackage{graphicx}
\usepackage[pdftex]{hyperref}
\idline{volNum}{2001} 

\begin{document}
	\def\teff{$T\rm_{eff }$}
	\def\kms{$\mathrm {km s}^{-1}$}
	
	\newcommand{\citeauthyear}[1]{\citeauthor{#1}, \citeyear{#1}}
	\newcommand{\citeauthinyear}[1]{\citeauthor{#1} in \citeyear{#1}}
	\newcommand{\urlnote}[1]{\footnote{\texttt{#1}}}
	
	\title   {The Virtual Observatory and Grid in Spain}
	\subtitle{VO and Grid Infrastructures and Initiatives}
	
	\author{
		J.D. \,Santander-Vela
	}
	
	\offprints{J.D. Santander-Vela}

	\institute{
		\email{jdsant@iaa.es}
		Instituto de Astrof\'isica de Andaluc\'ia  --
		CSIC, Apdo. Correos 3004, Granada 18080, Spain 
	}
	
	\authorrunning{J.D. Santander-Vela}
	\titlerunning{VO and Grid in Spain}

	\abstract{
		The Virtual Observatory (VO) is nearing maturity, and in Spain
        the Spanish VO (SVO) exists since June 2004. There have also
        been numerous attempts at providing more or less encompassing
        grid initiatives at the national level, and finally Spain has an
        official National Grid Initiative (NGI). In this talk we will
        show the VO and Grid development status of nationally funded
        initiatives in Spain, and we will hint at potential joint
        VO-Grid use-cases to be developed in Spain in the near future.
		
	}
	
	\maketitle
	
	\section{Introduction} 
	\label{sec:introduction}
	
	The Virtual Observatory, as proposed by
	\citeauthinyear{2001Sci...293.2037S}, is today a tangible 
	infrastructure used by thousands of astronomers (and 
	non-astronomers) every day. It help scientists by letting them 
	explore massive amounts of data, to derive statistical properties 
	from thousands of objects, and even to find new kinds of objects; 
	by providing a multi-wavelength view of particular objects or 
	regions of the sky; and by providing time-lapsed views of any 
	region of the sky from archived data.
	
	As mentioned, the foundational paper of the Virtual Observatory by
	Szalay and Gray was written in 2001, and the first Astronomical 
	Virtual Observatory (AVO) prototype, based on CDS's Aladin Sky
	Atlas\urlnote{http://aladin.u-strasbg.fr/} is from 2002 (see
	\citeauthyear{2005ASPC..347..183P}). By that time, our research team
	(the AMIGA group, Analysis of the interstellar Medium of Isolated
	GAlaxies, \citeauthyear{2005A&A...436..443V}) was working on a
	multi-wavelength study for the 1,051 galaxies in the Catalogue of
	Isolated galaxies by \citeauthinyear{1973SoSAO...8....3K} (see also 
	her online catalogue, \citeauthyear{1997yCat.7082....0K}). The 
	group had already compiled quite a large amount of information 
	(revised positions, IR fluxes in different wavelengths, images with 
	different filters), but still needed more information to be 
	retrieved, and wanted to make our different revised data products 
	available to the community.
	
	Using the existing International Virtual Observatory
	Alliance\urlnote{http://ivoa.net/} (IVOA) Proposed Recommendations,
	such as the VOTABLE (\citeauthyear{Ochsenbein:2004ty}) and the 
	Simple ConeSearch (\citeauthyear{Williams:2008fv}), the AMIGA group 
	finally provided a public web 
	site\urlnote{http://amiga.iaa.csic.es:8080/DATABASE/} delivering 
	data in VOTABLE format, providing at the same time a Simple 
	ConeSearch interface.
	
		\subsection{Radio astronomy and the VO} 
		\label{sub:radio_astronomy_and_the_vo}
		
		As astronomy started when the first humans raised their heads 
		and reckoned that the sky was very similar from night to night, 
		with seasonal variations that repeated from year to year, the 
		visible part of the spectrum has always been the most used, and 
		more familiar, to all astronomers. However, the discovery by 
		\citeauthinyear{Jansky:1933db} of a extraterrestrial and 
		extrasolar radiation with a wavelength of 14.6 meters, opened a 
		new spectral window that even today is just 63 years old, a 
		small fraction of the almost 400 years of instrumental optical 
		astronomy.
		
		The Virtual Observatory is also developing more rapidly for
		data from the visible parts of the electromagnetic spectrum, but 
		radio astronomical data is of the utmost importance for 
		understanding the most distant and obscured objects in the 
		universe, and the integration of these data sets in the VO 
		infrastructure has been a goal of our group since the beginning.
		
	
		\subsection{Grid and astronomy} 
		\label{sub:grid_and_astronomy}
		
		The computing grid was conceived by 
		\citeauthinyear{Foster:1999xe} as \emph{a hardware and software 
		infrastructure that provides dependable, consistent, pervasive 
		and inexpensive access to high-end computational capabilities}. 
		The similarity with the power grid arises from the fact that 
		until the power grid became common-place, electricity based
		facilities depended on their own power generation. Providing
		computing-on-demand, as easy to access as the power grid, is the 
		aim of grid computing initiatives.
		
		The next generation of astronomical instruments, that we might 
		call surveying instruments, such as the Large Synoptic Survey 
		Telescope (LSST), the Square Kilometre Array (SKA), the LOw 
		Frequency ARray (LOFAR), or even the Atacama Large Millimetre 
		and submillimetre Array (ALMA), will be extremely sensitive, 
		and at the same time their schedule will be completely 
		automated, so that the raw data rates will overwhelm
		any conventional computing facility. For these telescopes, 
		parallel processing of incoming data is needed, either by 
		extensive pre-processing at the data generation site, or by 
		running parallel pipelines for different levels of data 
		products, or by splitting the different levels of processing 
		between different \emph{tiers}. An extreme, present case, is 
		the LOFAR which, as can be seen from 
		\citeauthor{Valentijn:2008cr}'s proceeding, will need a super-
		computer class data reduction engine with a tiered architecture 
		for data reanalysis.
		
		At the same time, the storage needs for this instruments will 
		be so demanding that for some applications no actual data 
		access will be possible, but only streaming access to the 
		data being provided. Again, the LOFAR is an example of this. 
		In any case, data access has also to be distributed, and 
		different grid techniques (grid-FTP, IBM's Grid Parallel File 
		System...) are used.
		
	
	In the following sections, we will learn about the SVO (the Spanish
	Virtual Observatory), grid developments in Spain, and joint grid and
	virtual observatory efforts in our country. We will end with a 
	conclusions section.
	
	
	\section{VO in Spain: the SVO} 
	\label{sec:vo_in_spain}
	
	Publicly-funded VO activities in Spain are organised around the 
	Spanish Virtual Observatory (SVO\urlnote{http://svo.laeff.inta.es/},
	\citeauthyear{2006ASPC..351...19G}). Enrique Solano is the PI of the
	SVO, which joined IVOA in July 2004. With the creation of a publicly
	funded thematic network on the Virtual Observatory, the SVO has 
	spurred collaborations between all people with interest in the VO, 
	from scientists who wanted to use VO applications or technologies, 
	to data centres wishing to provide VO-compatible archives, with 
	groups wanting to publish data in the VO in between.
	
	Nowadays, the SVO provides 4 Full Time Employees (FTEs) from the
	LAEFF-INTA devoted to VO tasks, while the IAA-CSIC provides another 
	FTE. The LAEFF-INTA group constitutes what is known as the SVO-core, 
	and its continuity is guaranteed by recurring funding by the 
	Instituto Nacional de T\'ecnica Aeroespacial (INTA).
	
	The SVO-core has developed, and maintains, several VO archives, 
	such as the INES\urlnote{http://sdc.laeff.inta.es/ines/} (IUE 
	Newly Extracted Spectra), 
	GAUDI\urlnote{http://sdc.laeff.inta.es/gaudi/} (CoRoT
	Ground-based Asteroseismology Uniform Database Interface, see
	\citeauthyear{2005AJ....129..547S}), and
	OMC\urlnote{http://sdc.laeff.inta.es/omc/} (INTEGRAL mission Optical
	Monitoring Camera, see \citeauthyear{2004ASPC..314..153G}), among
	others.
	
	This experience will be key for the participation of the SVO-core 
	in the CONSOLIDER consortium for the GTC, both for the VO archive 
	of the telescope, but also for the artificial intelligence 
	techniques for scientific exploitation.
	
	\begin{figure}[tbp]
		\centering
			\includegraphics[width=\columnwidth]{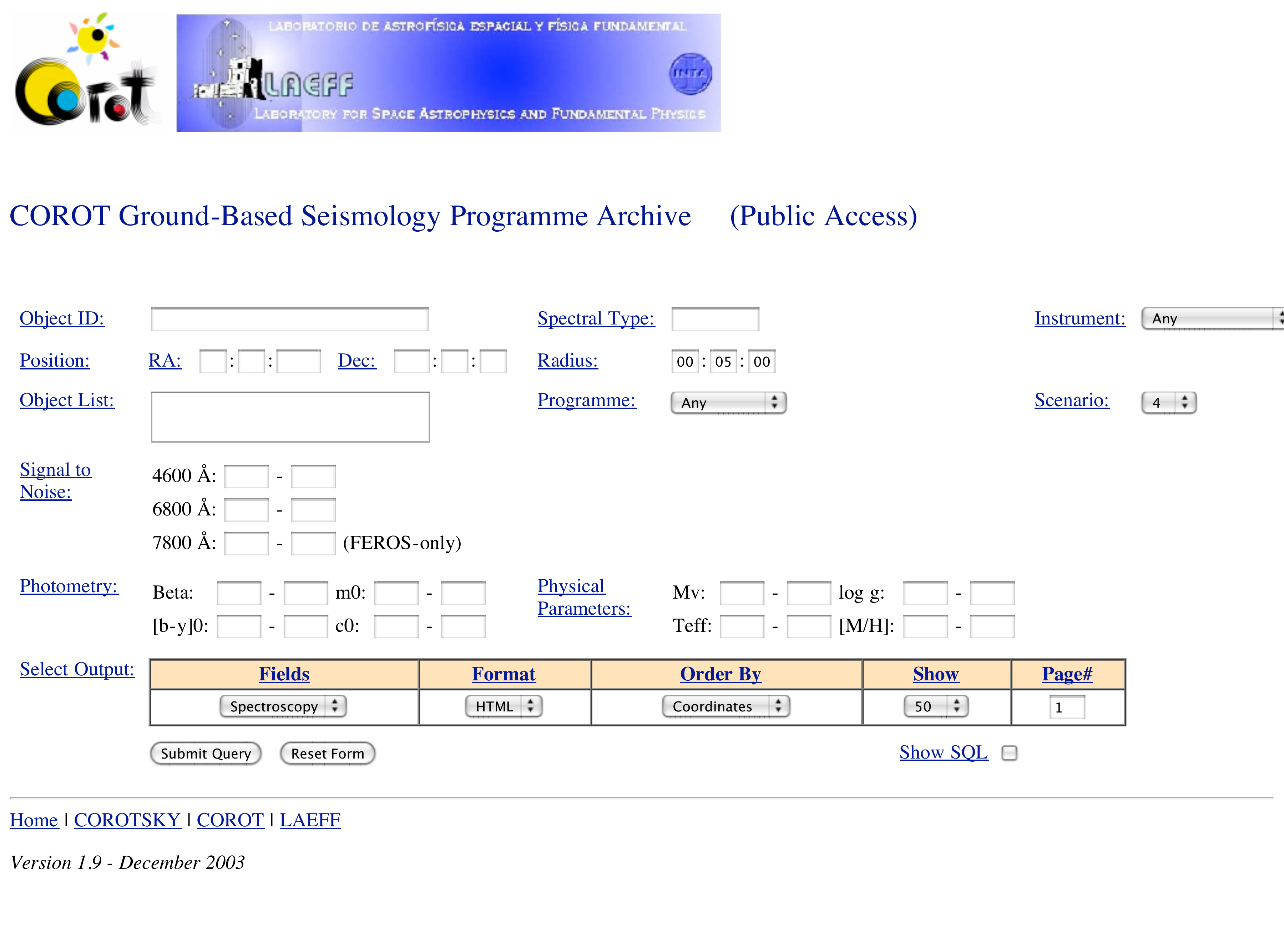}
		\caption{Main entry page for the VO-compliant GAUDI archive,
		         developed for the preparation of the CoRoT mission 
		         by the SVO-core.}
		\label{fig:CorotArchive}
	\end{figure}
	
	The SVO maintains presence in different VO-related international 
	bodies and projects: Enrique Solano is part of the IVOA Executive, 
	and the LAEFF-INTA is member of the VOTech, EuroVO-DCA, and 
	EuroVO-AIDA EU funded programmes.
	
\begin{figure}[tbp]
		\centering
			\includegraphics[width=\columnwidth]{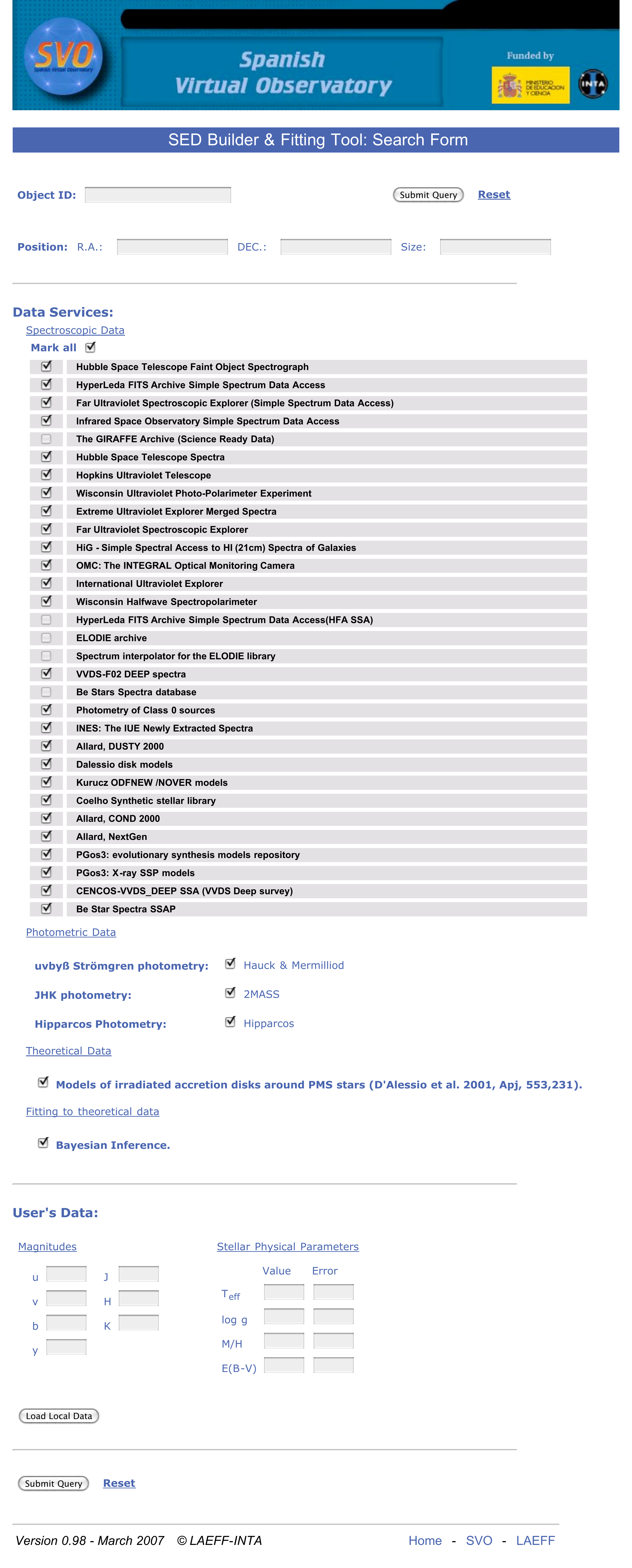}
		\caption{Web page of the VOSED search form. The user enters an 
		         object name, or coordinates, and selects the different
                 SSA services that can provide spectra for that
                 particular position. Photometry data can also be
                 retrieved to complete and create a synthetic SED, that
                 the user can fit with VOSED to different models.
		}
		\label{fig:vosed}
	\end{figure}

	Fostering VO-enabled science, science performed with VO tools, has
	always been one of the main concerns of the SVO. In that regard, two
	kinds of tools have been developed by the VO: ready to use web-based
	tools, such as the VOSED\urlnote{http://sdc.laeff.inta.es/vosed/} 
	(see figure~\ref{fig:vosed}), and artificial intelligence/data 
	mining tools.
	
	The VOSED provides an interface to Simple Spectral Access (SSA, see
	\citeauthyear{Tody:2007yq}), photometry, and synthetic spectra 
	services, and even to user provided spectra files, so that a user 
	can compile a multi-wavelength Spectral Energy Distribution (SED) 
	that can later be fitted to simple black-body emissions, or to more 
	complex star and dust/debris disks models.
	
	In order to access theoretical spectra, the SVO also developed the
	Theoretical Spectra Access Protocol (TSAP, see
	\citeauthyear{2007arXiv0711.2629R}), an extension to the SSA with
	additional parameters that allow applications to find out supported
	models, and for spectra to be synthesised on the fly from them. The
	VOSED uses the TSAP to perform the theoretical spectra fitting.
	Additional theoretical related efforts by SVO members include the
	PGos3\urlnote{http://ov.inaoep.mx/pgos3/index.php} theoretical model
	database for stellar populations, developed in coordination with the
	Mexican National Institute for Astrophysics, Optics and Electronics
	(INAOE).
	
	Data mining is another constituent of the SVO. The VO allows 
	access to large amounts of data, but being able to extract 
	meaningful properties from huge sets of objects, such as 
	classification in different kinds, or finding completely new kinds 
	of objects, is only possible by means of data mining techniques, 
	such as neural networks, Support Vector Machines, k-Means 
	algorithms... One example of neural networks use at
	the SVO is the paper on automated classification of eclipsing 
	binaries by \citeauthinyear{2006A&A...446..395S}.
	
\begin{figure}[tbp]
		\centering
			\includegraphics[width=\columnwidth]{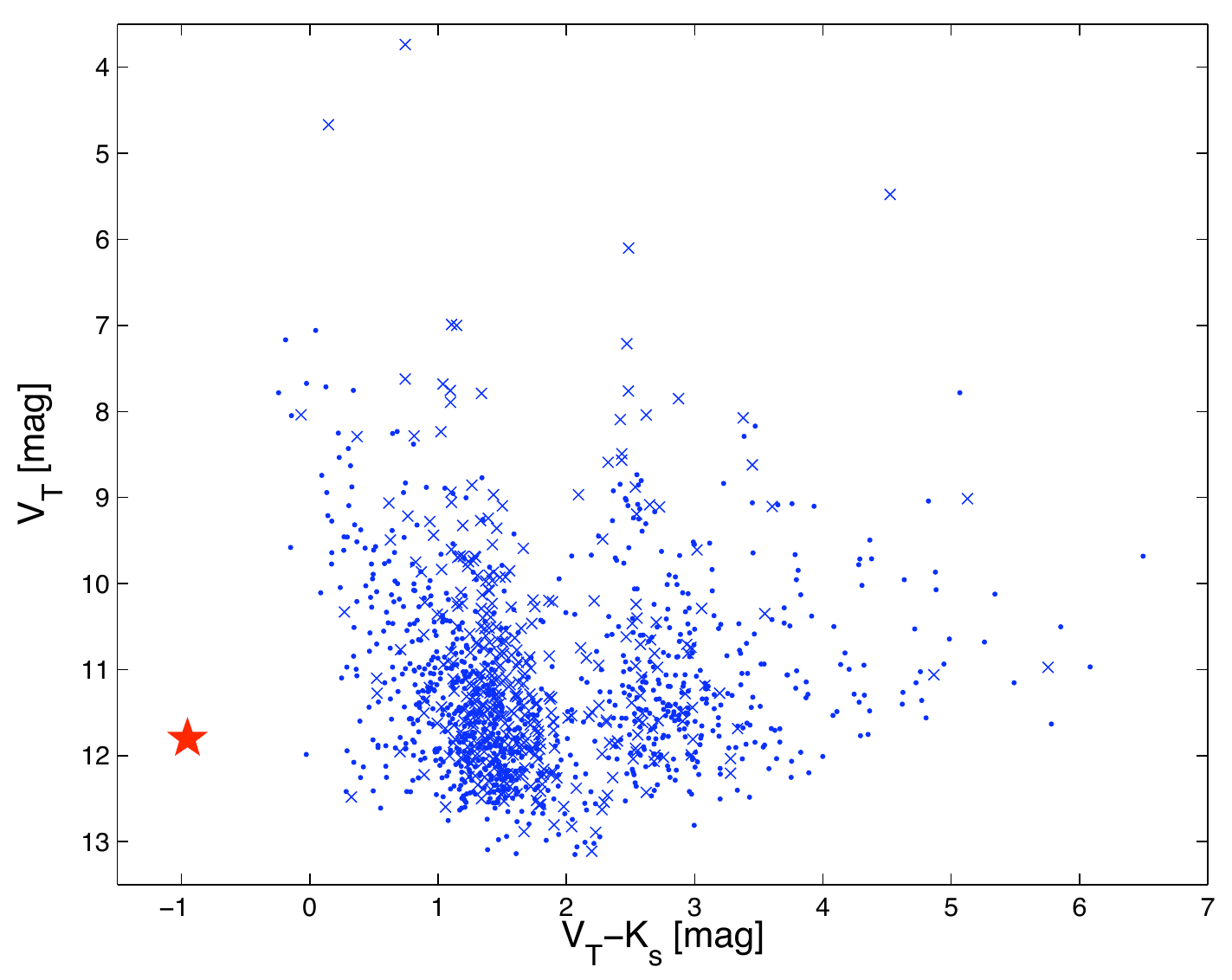}
		\caption{
		        $V_T$ versus $V_T - K_s$ color-magnitude diagram from
                the data in \cite{2008arXiv0804.2184C}. Tycho-2/2MASS
                sources with proper motions larger and smaller than
                15~mas~$\mathrm{yr^{-1}}$ are shown with crosses and
                dots, respectively. Albus~1 is highlighted with a big
                filled (red) star.
		        }
		\label{fig:albus}
	\end{figure}

	Several science papers have been published by members of the SVO, 
	or using tools developed at the SVO. Members of the SVO take part 
	in the VSOP project (\citeauthyear{2007A&A...470.1201D}) for 
	variability-type determination through data mining techniques for 
	not yet characterised objects, and several serendipitous 
	discoveries have been performed, such as that of Albus~1, a very 
	bright white dwarf candidate discovered by
	\citeauthinyear{2007arXiv0707.1343S}. In figure~\ref{fig:albus} we 
	can see Albus~1 is clearly an outlier in the Tycho-2/2MASS color
	distribution.
	
	\section{Grid in Spain} 
	\label{sec:grid_in_spain}
	
	Foster's definition of a grid can be posed as \emph{computing power 
	as easy to use as electricity}. Of course, that means that grid 
	computing is a subset of networked computing, and any grid computing 
	initiative is only meaningful when there is a powerful enough 
	network linking available computers.
	
\begin{figure}[tbp]
		\centering
			\includegraphics[width=\columnwidth]{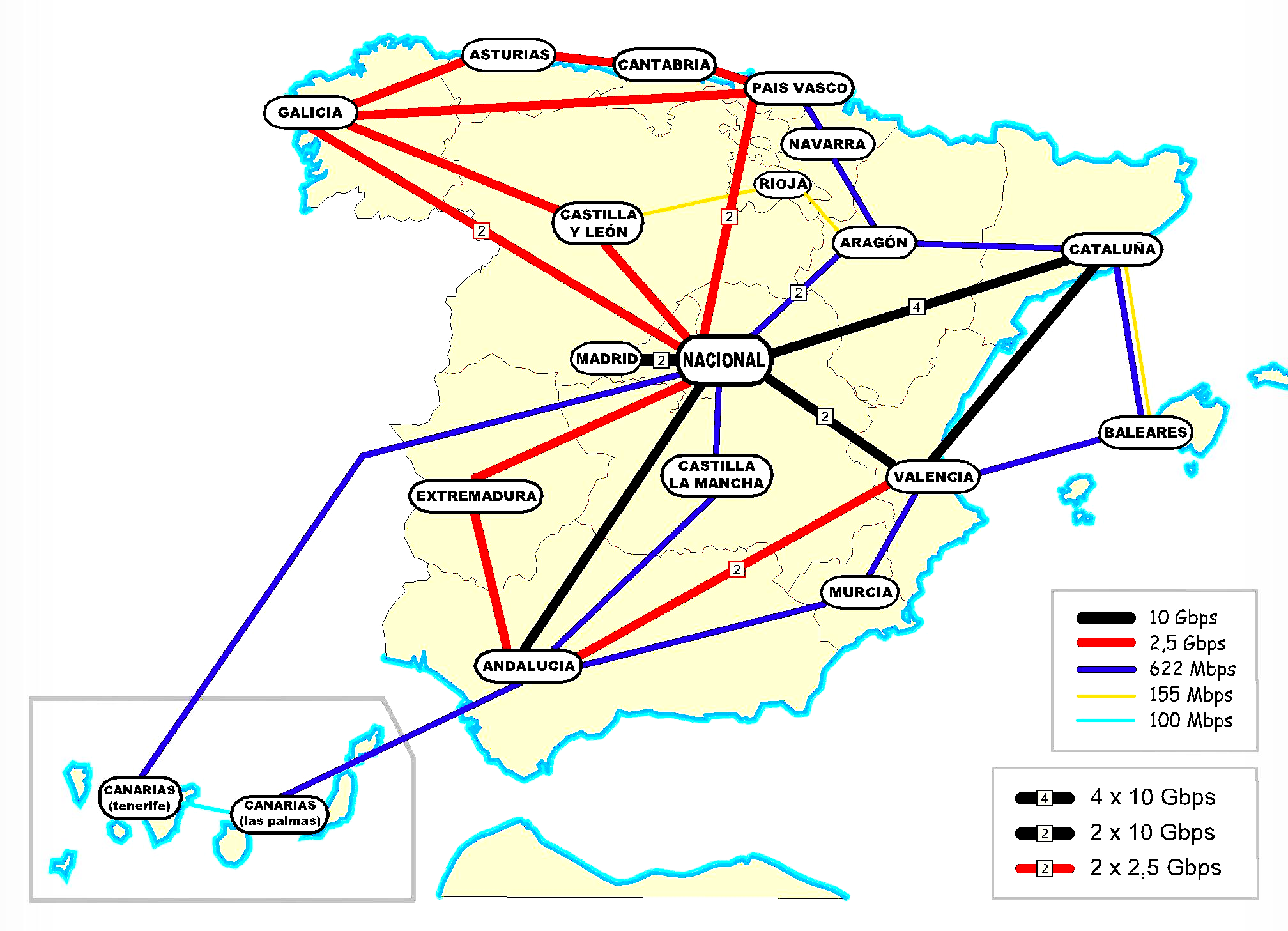}
		\caption{
				RedIRIS high-level topology. Each region has its own
                networking centre that provides service to capital
                cities, with county access relayed from them. RedIRIS is
                connected in Madrid to the G\'eant2 network, to the
                global internet, and to Spanish internet through the
                ESPANIX internet exchange point. Regional topology not
                shown.
		}
		\label{fig:RedIris}
	\end{figure}

	In Spain, all universities and research institutes are connected 
	via a high-bandwidth network infrastructure that was known as 
	RedIRIS (see figure~\ref{fig:RedIris}), and now is known as
	Red.Es\urlnote{http://www.red.es/}, a Spanish initiative for a 
	networked society. RedIRIS was connected to the European research 
	network G\'eant2 in March 2006, thus providing more bandwidth for 
	research interoperation in the EU.

\begin{figure}[tbp]
		\centering
			\includegraphics[width=\columnwidth]{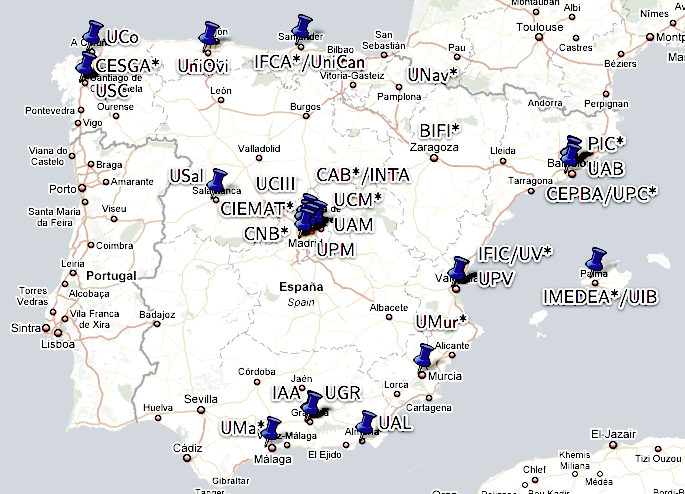}
		\caption{
				Map of the different research groups/institutes
                initially forming the IRISgrid initiative. Groups with
                an asterisk (*) provided computing nodes for the
                IRISgrid computing demonstration on December 2004.
		}
		\label{fig:IrisGrid}
	\end{figure}

	From the networking research aspect of RedIRIS, research groups 
	started to promote the need for a grid initiative, and thus 
	IRISgrid was born in 2003. Several research groups had also gained 
	experience by participating in the Enabling Grids for E-sciencE 
	(EGEE), CrossGrid, EU-DataGrid, Interactive European Grid 
	(Int.eu.grid, also known as I2G) and other European-level grid 
	projects. The teams involved in IRISgrid are shown in 
	figure~\ref{fig:IrisGrid}.
	
	Not surprisingly, most of the centres participating in IRISgrid come
	from the High Energy Physics/Particle Physics centres. Spanish
	experience also includes grid middleware development: the Department 
	of Computer Architecture and Automation 
	(DACYA\urlnote{http://www.ucm.es/info/dacya/}) at the Universidad
	Complutense de Madrid, has developed the
	GridWay\urlnote{http://www.gridway.org/} meta-scheduler, now part of
	EGEE's gLite\urlnote{http://glite.web.cern.ch/glite/} middleware
	distribution.
	
	However, funding for the IRISgrid initiative ended in the beginning 
	of 2007. Moreover, IRISgrid was never officially recognised by the 
	Spanish ministry of Science and Education as THE national grid 
	initiative. When the European Grid 
	Initiative\urlnote{http://web.eu-egi.org/} started to
	take shape in 2007, prompting possible members to register with 
	their own initiatives, finally an officially recognised grid 
	initiative was launched: the Spanish Network for
	e-Science\urlnote{http://www.e-ciencia.es/}.
	
	Still, the Spanish Network for e-Science consists mostly of the 
	centres participating in the IRISgrid initiative, plus centres who 
	wish to learn and start using grid techniques, but no hardware or 
	budget (aside from continued funding for the academic 
	inter-networking through Red.es) has been provided.
	
	In this context, and taking into account that many of the research
	centres with extensive grid experience belong to the CSIC Spanish
	Research Agency (Consejo Superior de Investigaciones Científicas), 
	the GRID-CSIC initiative has been born this year
	(\citeauthyear{Marco:2008qv}) with the aim to provide CSIC centres 
	with high-end computing facilities that will be networked and 
	provided with EGEE- and I2G-compatible middleware so that they can 
	be seamlessly integrated in the Spanish NGI.
	
\begin{figure}[tbp]
		\centering
			\includegraphics[width=\columnwidth]{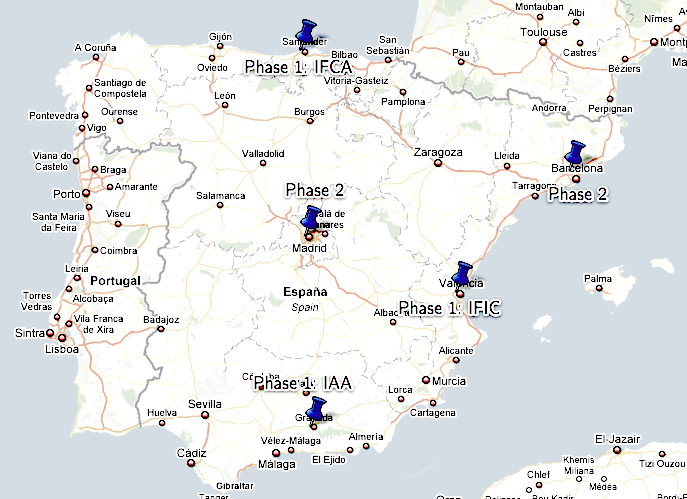}
		\caption{
				Centres participating in the two initial phases of
                GRID-CSIC. The IFCA and IFIC belong to the high-energy
                physics community, and IAA is the Institute for
                Astrophysics of Andalusia, and uses grid and super
                computing for planetary atmospheres, stellar structure
                modelling, and N-body simulations, among others. The
                second phase will include additional, yet to be defined,
                nodes at Madrid and Catalonia.
		}
		\label{fig:GRID-CSIC}
	\end{figure}

	This infrastructure must be compatible, and eventually shared with 
	the joint Spain-Portugal IberGrid initiative, and with the France 
	CNRS Institut des Grilles (Grids Institute). It will be started in 
	2008-2009 with a test phase with initial deployment of just three 
	nodes at centres with grid experience (IFCA, IFIC, IAA), with three 
	yet to be defined additional nodes in Madrid and Catalonia for 
	2009-2010, and the rest of the nodes operational in 2010 (see 
	figure~\ref{fig:GRID-CSIC}). Total estimated computing power at 
	the end of the project will be in the order of 8000 cores (either 
	AMD Barcelona or Intel Xeon x86\_64 architecture
	processors), with on-line storage of up to 1 PB (1000 TB).
	
	Apart from project coordination and infrastructure management, an
	additional application development and grid porting support area has
	been defined, so that users can make real use of the GRID-CSIC
	equipment. One of the most relevant evaluation criteria of the 
	project success will be the percentage of users making active use 
	of the grid for their research.
	
	IAA experience in IrisGRID comes from high-performance computing 
	users in the area of astrophysics, and was assigned with the task 
	of providing the use cases and background for IrisGRID in the topic 
	of astrophysics, while the IFCA and IFIC are Tier-2 centres for the 
	processing of LHC data. The IAA plays also a key role in the 
	e-Science for Andalusia regional e-science initiative, 
	e-CA\urlnote{http://e-ca.iaa.es/}.
	
	
	\section{The intersection of VO and grid in Spain} 
	\label{sec:vo_and_grid_in_spain}
	
	The VO provides an infrastructure analogous to the Data Grid 
	envisioned by \citeauthor{Chervenak:2001rr}, and thus allows for 
	parallel access from different nodes at the same time. Grid 
	computing within the VO needs to exploit the parallelism in 
	different use cases:
	
	\begin{description}
		\item[\textbf{Parameter space exploration:}]
				several astrophysical simulations, such as stellar
                structure, evolutionary tracks, et cetera, start running
                with a particular set of initial parameters, but many
                runs are required to sample a particular parameter
                space. In this case, the \texttt{gridway} meta-scheduler
                can be used to easily gridify this kind of applications,
                running instances of the same program that only differ
                in the input files/parameters. Many Monte Carlo
                simulations fall also in this category.
				
		\item[\textbf{Exploration of partitioned data sets:}]
				this case is very similar to the one above, but the
                parameter space is partitioned instead of sampled,
                allowing instances to run independently of each other on
                each partition. Examples of this are massive object
                searches for particular kinds of objects, which might
                need querying large data sets, even from many different
                archives, but the search for properties of those objects
                can be made independently of each region of the sky.
                This partition is not necessarily spatial: data
                reduction pipeline tasks partition data based on
                observation blocks.
				
		\item[\textbf{Loosely coupled simulations:}]
				the above mentioned cases work better because each node
                in the grid does not have to communicate with others, or
                communication is restricted at the end and the beginning
                of the process. N-body simulations tend to be just in
                the opposite corner: tracking all bodies tends to be
                more communications-bound than processing time bound.
                However, techniques such as Smoothed Particle
                Hydrodynamics, and hierarchical N-body simulations are
                more grid friendly, by restricting interaction between
                grid elements to high-level interaction with the
                following level in a hierarchy.
				
	\end{description}
	
	Many applications cannot be so easily partitioned. One example are
	searches of related bodies (pairs or triplets of galaxies, for
	instance): the partitioning technique can be used, but either the
	partitioning needs overlapping, or special border cases have to be
	considered, reducing efficiency in any case.
	
	In the above discussion of astrophysical applications of the grid we
	have not mentioned the Virtual Observatory. There are different 
	aspects of the VO where grid computing might be beneficial:
	
	\begin{description}
		\item[\textbf{Visualisation:}]
				most of the VO visualisation data comes directly from
                web services. However, some of those images can be
                generated on-the-fly by powerful enough systems. If the
                \emph{ad-hoc} generation tools are gridified, the system
                would make transparent use of the grid infrastructure
                without passing the complexity of the system to the
                user.
				
		\item[\textbf{Data access:}]
				VO protocols such as the VOSpace
                (\citeauthyear{Graham:2008hb}) for always available
                storage can be mapped on top of the Data Grid by means
                of implementations that make use of protocols such as
                Grid-FTP, IBM's Global Parallel FileSystem, et cetera.
                However, access policies still hinder the adoption of
                these protocols.
				
		\item[\textbf{Data processing:}]
				the VO paradigm is to perform analysis as near to the
                data as possible, in order to reduce network
                bottlenecks. A pervasive grid computing facility makes
                it easier for data providers to allow for sophisticated
                analysis tools running against the data, without having
                to worry about computing resources and scalability.
                There is an IVOA Note on the Common Execution
                Architecture (\citeauthyear{Harrison:2005la}), an API
                for making VO aware analysis tools available as web
                services. This services are the facade client
                applications see, and again, the grid can be used in a
                transparent way.
		
		\item[\textbf{Data mining:}]
				given the rich metadata available for VO-compatible data
                sets, data mining applications are particularly well
                suited to the VO. Additionally, more and more algorithms
                are being deployed in the form of web-services, what
                helps in the development of a distributed infrastructure
                for knowledge extraction.
		
	\end{description}
	
	Most of grid experience in Spain, even in the field of astrophysics, 
	has not been connected with the Virtual Observatory in any way, but 
	more and more users are planning to make use of it. High level data 
	analysis techniques are being prepared by the AMIGA group for 3D 
	data sets, jointly with the SVO-core, that will make use of the 
	grid. Groups at IAA working on stellar structure and evolution are 
	starting to their code bases aware of the grid within the e-CA 
	project framework, and many data mining projects, such as VSOP, will 
	make use of the grid for their processing. And the Spanish members 
	of the IVOA Theory Interest Group, as part of the SVO, wish to use 
	VO tools, and keep developing access protocols such as TSAP, for 
	“micro” simulations, i.e. not cosmological simulation, but 
	special interest simulations, such as stellar structures, initial 
	mass function distributions, et cetera.
	
	
	\section{Conclusions} 
	\label{sec:conclusions}
	
	We have shown that pure Virtual Observatory activities in Spain 
	have a very good health, and there is ample experience both inside 
	and (specially) outside the astrophysics community, but more and 
	more services will be deployed on the grid as it becomes 
	more pervasive. Spain will have in the near future a mature enough 
	grid infrastructure within the European Grid Initiative, and 
	different research groups are already porting, or planning to port,
	their code bases to the grid, taking as much as possible from 
	existing VO infrastructures.


	\begin{acknowledgements}
		This work has been partially funded by the SVO network 
		(MEC AYA2005-24102-E project) and the EuroVO-DCA FP6 program. 
		It has also been partially funded by the MEC AYA 2005-07516-C02 
		project. The author also wishes to acknowledge inputs from SVO's 
		head, Enrique Solano, and Jos\'e Ruedas, director of the computing 
		facilities at IAA-CSIC.
	\end{acknowledgements}
	
	\bibliographystyle{aa}
	\bibliography{ngi_vo_spain}

\end{document}